# Evaluation of silicon consumption for a connectionless Network-on-Chip

Marcelo Daniel Berejuck
Electrical Engineering Department
University of Southern Santa Catarina - UNISUL
Florianópolis-SC, Brazil
marcelo.berejuck@ieee.org

Antônio Augusto Fröhlich
Computer Science Department
Federal University of Santa Catarina - UFSC
Florianópolis-SC, Brazil
guto@ieee.org

*Abstract* — We present the design and evaluation of a predictable Network-on-Chip (NoC) to interconnect processing units running multimedia applications with variable-bit-rate. The design is based on a connectionless strategy in which flits from different communication flows are interleaved in the same communication channel between routers. Each flit carries routing information used by routers to perform arbitration and scheduling of the corresponding output communication channel. Analytic comparisons show that our approach keeps average latency lower than a network based on resource reservation, when both networks are working over 80% of offered load. We also evaluate the proposed NoC on FPGA and ASIC technologies to understand the trade-off due to our approach, in terms of silicon consumption.

*Network-on-Chip; System-on-Chip; Quality-of-Service; Multimedia*

## I. Introduction

The silicon industry has used Systems-on-Chip (SoC) with multiple heterogeneous processing units as means to deliver the performance required by modern multimedia applications [1]. However, the integration of an increasing number of specialized processing units poses a challenge on the interconnection mechanisms in such systems. These systems are now required to handle a large number of very distinctive communication flows, with very distinct latency and bandwidth requirements. As a solution, the silicon industry has been using predictable Networks-on-Chip (NoC) to interconnect components in this kind of SoC.

Such NoCs must allow guarantees on bandwidth and latency to be made for individual flow in the network, and end-to-end connections are suitable solutions to provide these guarantees. There are two known techniques to achieve this: circuit switching and non-blocking routers with rate control. SoCBUS [2][3] and 4S [3][4][5] are examples of NoCs based on circuit switching. On this technique, when the connections are established, they do not share resources until the end of packet transmission; thereby, real-time guarantees are easily achieved. Time Division Multiplexing (TDM) is another technique of circuit switching. It is employed in networks such as Æthereal [7][8], dAElite [7] and Nostrum [10][11], in which the time domain is divided into several recurrent time slots of fixed length, one for each channel. TDM implements virtual circuit switching, which may result in better resource utilization than pure circuit switching technique. As alternative for circuit switching, other networks adopt non-blocking routers with rate control as can be seen in MANGO [12][13][14]. On that approach, packets have arbitrated locally at switches and buffering and rate control are required in order to achieve real-time guarantees.

Nevertheless, after over a decade developing industrial Telecom and multimedia projects, we realized that many applications in this domain would profit better from a NoC that could optimize the utilization of resources for multimedia flows that tolerate reasonable variations in the Quality-of-Service (QoS). Indeed, most of the projects we worked on have conceived around a few very strict real-time communication flows, and a large number of less strict multimedia flows that tolerate much larger variations in latency and bandwidth.

In this context, this paper introduces a Network-on-Chip with worst-case latency (WCL) predictable at design time, in which there is no resources reservation. The proposed NoC architecture has based on the interleaving of flits from different flows in the same communication channel between routers, so each flit carries along routing information. As shown later, the interleaving of flits reduces the average latency for small packets and increases for bigger ones. To deal with that issue, we have reduced the number of hops in the network, increasing the number of processing elements on each router in the network. Usually, the routers for NoCs have with five communication channels: four channels for the connection with other routers in the network and one to connect the processing element. The router proposed in this paper has eight communication channels, and the trade-off due to this approach will be discussed in this document in terms of silicon consumption.

The remainder of this paper has the following organization: in Section II introduces the network concept describing the internal structure of the communications channels and router architecture proposed to build the NoC. Section III introduces a definition for latency analysis in NoCs and discusses the latency for the proposed network and a generic best-effort NoC. Section IV presents the experimental results of silicon consumption for FPGA and ASIC technologies, and Section V finalizes the paper with our conclusion.

## II. NETWORK CONCEPT

The network proposed in this paper has conceived for variable-bit-rate scenarios and therefore strategies based on resource reservation were ruled out in favour of deterministic scheduling. The basic idea is to embody each flit with routing and scheduling information, so that routing is performed flit-by-flit based solely on information locally available at each router in a way that preserves the determinism of the worst-case latency for each path. The additional overhead of carrying routing information along with each flit will be evaluated in Subsection II.F.

### A. Assumptions for the Network

We understand that the adoption of a routing algorithm and an arbiter that allows alternately access to the communication channel of a router ensure predictability for all flows in the network without reservation of network resources. Thereby we focused on finding techniques in order to turn it feasible. To achieve that goal, the following assumptions have taken into account:

- the routing is done flit-by-flit and made fairly among the flows that are competing for a communication channel;
- arbiters must grant priority to flits coming from distant routers;
- to minimize the competition for communication channels between routers, up to eight communication channels are available for each router, in order to explore the sense of locality; and
- once the routing is done flit-by-flit, the buffers are placed only on the end points, minimizing the side effect of growing on silicon area.

Next Subsections introduce the network architecture adopted to cover these assumptions.

### B. Network topology

The routers were conceived to be connected to form a 2-D orthogonal topology. Each router can be configured at design-time to have different communication channels quantities, from five up to eight as shown in Figure 1-a. The communication channels were named with cardinal points and can be connected either to one core or to another router in order to build larger networks.

We understand that the complexity of some elements in a router grows exponentially with the number of communication channels, but we had empirical evidence that locality plays a major role in real applications. Therefore, we decided to support up to eight ports per router, thus enabling designers to connect cores that will communicate more intensively with each other on the same router. Figure 1-b depicts a network with four routers and twenty-four processing elements connected in it.

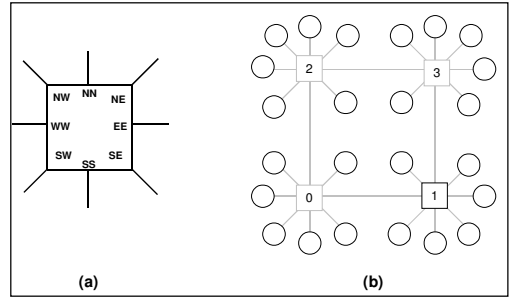

Figure 1. Network topology: (a) router with eight ports; and (b) example of network with four routers and twenty-four processing cores.

### C. Physical communication channels

Each router interconnection point provides two unidirectional channels, one for input and one for output, as shown in Figure 2. In addition to data (DIN and DOUT in the Figure) and address signals (AIN and AOUT in the Figure), each physical channel features control signals to synchronize the data transfer. The width of each channel can be configured at design-time based on the application's needs. Data words DIN and DOUT can have arbitrary width though typical applications will seldom deviate from the traditional 8–64 range. The address words AIN and AOUT are 2p+3 bits long: p bits for each of X and Y coordinates, with p designating the size of the network (for example, p=1 is a 2x2 network), and 3 bits for the local port H.

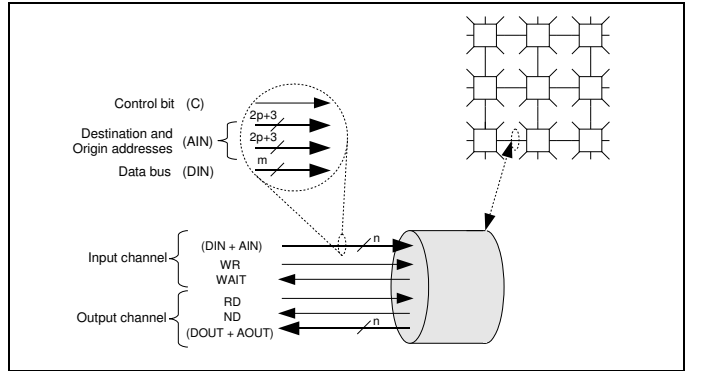

Figure 2. Internal structure of communication channels.

The control signals RD and WR are strobes used respectively to read data from the output channel and to write data into the input channel. The signals WAIT and ND are used for flow control. WAIT must be cleared before new data can be written into an input channel, and ND is set to indicate that a new flit is available to be read from the output channel.

### D. Packet format

The structure of the physical channel inherently defines the format of a flit in the network, which is depicted in Figure 3. Each flit has 1+2(2p+3) + d bits, where p is the size of the network and d is the width of the data word (DATA in the figure). The first bit (C in the figure) is a control bit to distinguish a data flit from a control one. The adjacent bits are for addressing: $X_{ORI}$ and $Y_{ORI}$ are the X and Y coordinates of origin router and $H_{ORI}$ designates the port in that router from which the flit was injected in the network (equivalent to a host on a cluster, therefore H); $X_{DST}$, $Y_{DST}$ and $H_{DST}$ carry the counterpart destination information. The remaining bits are for data.

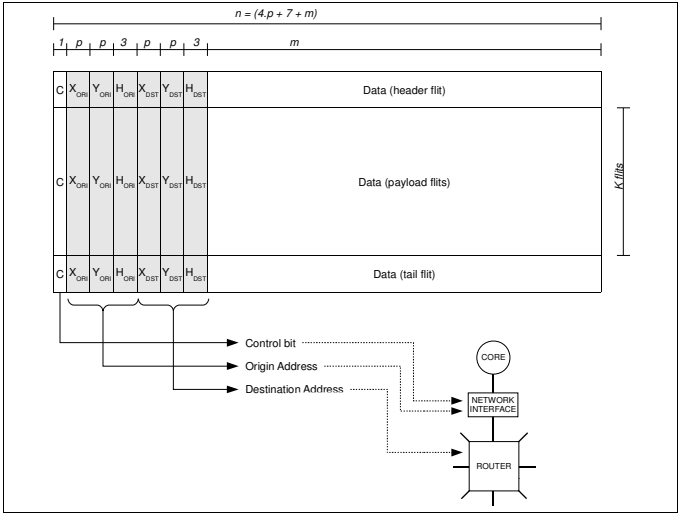

Figure 3. Network's packet format. The size of X/Y fields depends on the network size.

From the point of view of its routers, the network's packet is an arbitrary-length sequence of payload flits preceded by a header flit and followed by a tail flit. Header and tail flits are marked as control flits by setting the control bit field (C in Figure 3), while payload flits are marked as data by having the same bit cleared. That information placed in the field DATA field of a header or tail flit is seen by routers as higher-level protocol control information and therefore is not handled by them. From the point of view of the network adapter, however, the packet depicted in Figure 3 has a maximum size specified at design-time due to the design of buffers in the network interface. This point will be discussed in Subsection G. Figure 3 also depicts the structures (network components) in the network that handle each of fields above described. The fields Control Bit (C) and Data are used by an interface called Core interface, while the router uses Destination Address.

*E. Router components*

Figure 4 depicts the internal structure of a router. It was implemented with seven logic blocks (or components): input interface, output interface, flow control, routing control, arbiter control, allocator, and crossbar switch. These components will be described in the following Subsections, as well as the synchronization among them.

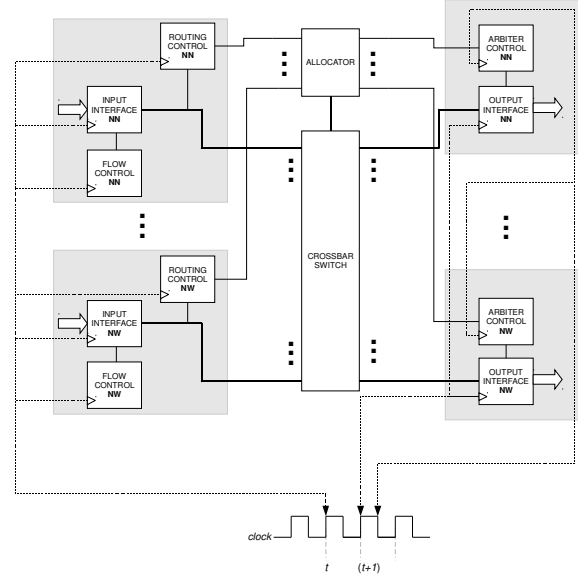

Figure 4. Block diagram showing the internal structure of the router.

*1) Input interface and flow controller*

The input interface shown in Figure 4 is a block that receives flits from the communication channel such as shown in Figure 2. It has a register that may store a flit. After receive a flit into its register, the input interface informs another block called flow controller that there is a new flit waiting for routing grant.

The flow controller checks the destination fields from the flit stored in the input interface register and informs to another block called routing controller about the destination requested in the flit. Meanwhile, the input interface sets the WAIT signal, shown in Figure 2 while the received flit still waiting for its routing. The signal WAIT will be turn off after the input interface receive a command from the destination output in the router, informing that it is available to receive another flit. This information comes from the Arbiter Control that will be describer later in this Section.

*2) Allocator and Crosbar Switch*

An n x m Allocator is a unit that accepts n m-bit vectors as input and generates n m-bit vectors. The Allocator of router has eight three-bit vectors as input and eight three-bit vectors as output. It uses its input interfaces to receive the commands from the Arbiters and control the Crossbar Switch, using its output interfaces to perform the connections requested by the Arbiters. The Allocator works in parallel mode in order to improve the switching performance of the Crossbar. It means that, in one clock cycle, it can receive up to eight commands from the Arbiters and send up to eight commands to the Crossbar.

An n x m Crossbar Switch is a structure that directly connects n inputs to m outputs with no intermediate stages. In effect, such as a switch consists of m n:1 multiplexers, one for each output, the Crossbar Switch of router are a square Crossbar with eight inputs and eight outputs. It was configured to connect any input channel of the router to any output channel. However, it can be done under the constraint that each input is connected to at most one output, and each output is connected to at most one input, as well. The structure with multiplexers adopted ensures that the constraint that each input is connected to at most one output will be accomplished. Furthermore, the Crossbar Switch does not use clock cycles to perform the switching, and each output may receive data from one of those inputs at the same time.

*3) Output Interface*

The output interface is a block that receives flits from the Crossbar Switch. It may store one flit, such as happen to the input interface. After receive a new flit, it set the signal ND, shown in Figure 2, to inform the next router (or communication channel) that a new flit is available to be delivered. The signal ND will turn off after a reading in its register. It will be signaling by the RD signal, also shown in Figure 2. After that reading, the output interface informs to the input interfaces that it is available for new transmissions.

*4) Router latency*

The latency to forward a flit from an input channel to an output channel is two clock cycles. Figure 3 shown the synchronization process of internal components. For simplicity, Figure 3 depicts only one input and output interfaces for North (NN) and Northwest (NW) communication channels.

There is a clock signal on the bottom of Figure 3. Note that, at time t, the three components of input channel are active by the rising edge of the clock. On the next cycle, (t+1), the output interface notifies that a new flit is available at the output channel. Considering that the reading of output channel is done at the same time that the notification has occurred, then the flow controller is notified about reading and the arbiter is updated at the next falling edge of the clock. Thereby, a flit can be forwarded at two clock cycles.

*5) Routing algorithm*

A key element of the design is flit-by-flit routing. Since every flit carries along its destination address, arbitration can be implemented locally on each router for each of its output ports. If there are several packets being routed through the same link, the arbiter will alternate access to the corresponding output port so that each flow gets to forward one flit at a time. Hence, the flits from different packets are interleaved in the network. Conversely, circuit-based or wormhole-routing networks would block the output port at least until the end of a packet. Additionally, the flit-level, interleave of flits routing strategy we have adopted largely simplifies buffering: routers only have to implement a single-flit buffer for each output port.

Routing of flits are performed using the XY routing algorithm, which ensures in-order, deadlock-free delivery for 2-D orthogonal networks [15][16]. Since there is only one routing path for the communication between any two cores in the network (note that flits are always routed first in the X axis and subsequently in the Y axis), the flits from a packet are delivered at the destination in the same order that they have been injected at the origin. The XY routing algorithm implemented in the routing control is shown in Algorithm 1.

---
**Algorithm 1**: XY

**Input**: $X_{DST}$, $Y_{DST}$
**Output**: Request

**if** ($X_{DST} = X_{ROUTER}$) **and** ($X_{DST} = X_{ROUTER}$) **then**
    Check $H_{DST}$ for local channel
    Request local channel
**else**
    **if** ($X_{DST} \neq X_{ROUTER}$) **then**
        **if** ($X_{DST} > X_{ROUTER}$) **then**
            Request East channel
        **else**
            Request West channel
        **end if**
    **elsif** ($Y_{DST} \neq Y_{ROUTER}$) **then**
        **if** ($Y_{DST} > Y_{ROUTER}$) **then**
            Request North channel
        **else**
            Request South channel
        **end if**
    **end if**
**end if**

---

*6) Arbiter algorithm*

Each output channel has its own arbiter to receive and manage the requests generated by the routing controllers at the input channels. The arbiter implements a scheduling algorithm, as shown in Algorithm 2.

Each channel receives a different priority level when the system starts. The highest priorities are given to the channels NN, SS, EE, and WW, since they are used to interconnect other routers in a 2-D regular mesh network. Furthermore, these channels can send more than one flit at each received grant. The amount of flits they can send at each grant depends on the number of requests that may occur at the same time, on the prior routers in the routing path.

---
**Algorithm 2**: Arbiter

**Input**: Request, Destination
**Output**: Grant, Channel

**if** (Reset = active) **then**
    Channels priorities ← initial values
    Channels counters ← initial values
**else**
    **if** (Request ≠ null) **then**
        Check priorities
        **wait until** (Destination ≠ *busy*)
        Forward flit
        Notify origin granted
        **if** (Request = *North, East, South* or *West*) **then**
            **if** (channel conter > 0) **then**

```
            Decrements channel counter
        else
            Channel counter ← initial value
            Channel granted ← lower priority
            Update priorities for other channels
        end if
    else
        Channel granted ← lower priority
        Update priorities for other channels
    end if
  end if
end if
```

Any flit has its routing request attended if it has the highest priority, or if there are no other requests in the arbiter. Once the request is attended, the channel that requested the sending of a flit receives the lowest routing priority level and may only send other flits if there is no other flit waiting for routing. Once the channel that has the routing priority is chosen, the arbiter sends a command to a logical block called Allocator, informing which routing has to be executed in that output channel

Since the arbiter allows alternate access to the router output port, there is no contention of output channels. Hence, flits from different packets are interleaved in the channels of the network. The adoption of XY routing and this arbiter ensure predictability for all flows in the network without reservation of network resources.

*F. Buffers in Network Interface*

The network interface (NI) is composed by two FIFO memories, one logic block to interface with the network, called "Router Adapter", and a logic block to interface with the processing unit (or core), called "Core Adapter". The internal structure of the NI is shown in Figure 5.

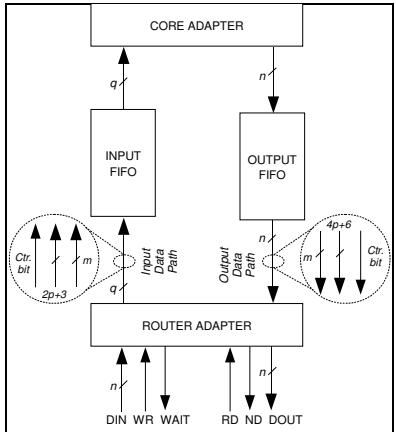

Figure 5. Internal structure of Network Interface.

The Router Adapter is a logic block that interacts with the network dealing with the signals of physical channel, introduced in Subsection C. It prepares the data that come from the network, to be delivered to the core, removing the Destination Address fields from the flit. The Destination Address is an information field used only by the routers in the network.

The Core Adapter also is a logic block that interacts with the core, and prepares the data that come from the core, to be written in the network, concatenating the fields Control bit, Origin Address and Destination Address to the Data field to generate the flit.

There are two FIFO memories in the Network Interface. One of them is used to store the data received from the network, and it is called Input FIFO while another one is used to store the data received from the core and it is called Output FIFO.

In the proposed NoC the throughput depends on the FIFOs sizes, which can be defined by the NoC designer. The router needs at least one flit stored and ready to transfer in the FIFO, in order to have efficient throughput. The latency between any two points in the network is not constant because the NoC's router does not reserve resources. It means that the FIFOs must be designed based on two factors. First, the time that the core spent writing/reading a flit in the network interface. Second, the lower value of latency expected on its communication with other cores through the network. The equation 1 describes that relationship between these times:

$$B_{size} = \left\lceil \frac{T_{wr/rd}}{T_{net}} \right\rceil \quad (1)$$

in which $B_{size}$ is the buffer size, $T_{wr/rd}$ is the time that the core spent to write/read a flit in the network interface, and $T_{net}$ is the lower value of latency expected to any core's communication through the NoC. $B_{size}$ is a parameter and must be defined before the synthesis of the network. When the value of $B_{size}$ is one, registers are synthesized, instead FIFO memories. Based on that premise, it is expected to have one flit per cycle as throughput in the network. The FIFOs have handshake signals to warning th cores when the memories are empty or full of flits. In fact, these handshaking signals (empty and full) are used as end-to-end flow control, and hence the size of FIFOs must be enough to store the flits and avoid the memory full state.

## III. LATENCY ANALYSIS

*A. Latency basics*

The latency of the Network-on-Chip is the time required for a packet to traverse the network, from the time the header of the packet arrives at the input channel to the time the tail of the packet departs the output channel [17].

The latency can be separated into two components:

$$L = T_h + \left( F/b \right) \quad (2)$$

in which $L$ is packet latency, $T_h$ is the time required for the header of the packet to traverse the network and $F/b$ is the time for the packet of length $F$ to cross a channel with bandwidth $b$.

In absence of contention, header latency might be seen as the sum of two factors determined by the topology: router delay and number of routers in a path between origin and destination. Based on these two factors, the equation (2) can be re-written as follows:

$$L = (H_{path} \cdot t_r) + (F/b) \quad (3)$$

in which $H_{path}$ is the number of hopes in the path and $t_r$ is the router delay. For simplicity, we do not include in the equation (1) and (2) the wire delay across the physical channel, even not the distance from the source and destination of a packet.

### B. Latency using Interleaving of flits

Let's suppose that there are three requests to send packets through the same communication channel at instant $t_0$, as shown in Figure 6-a, and the sequence of scheduling for those requests on instant $t_0$ is {Packet 1, Packet 2, Packet 3}. We define "Interleave of flits" the method in which the packets from all requests are broken into flits, and these flits have been sending through the channel, one flit from each packet at a time. The interleave of flits for this example is shown in Figure 6-b and a wormhole switching of those packets is shown in Figure 6-a

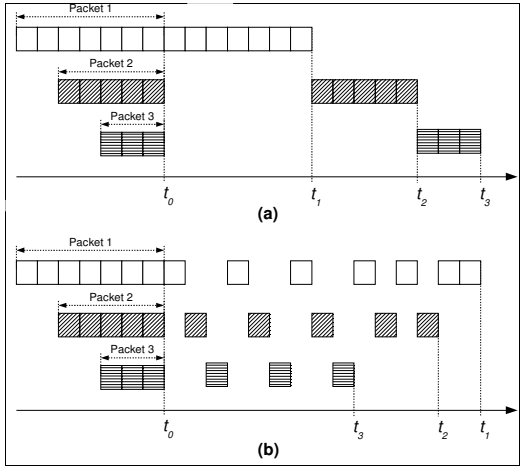

Figure 6. Example of interleave of flits in the same channels.

Note that the time to traverse the channel for the Packet3 using interleave of flits is ($t_3$- $t_0$), which is lower than the time spent in the wormhole switching. It means that the smaller size packets have better average latency when the method of interleave of flits is used, while bigger size packets have their average latency increased. This is a trade-off when the interleaving of flits is adopted. The method is suitable for the systems on which short packets must have their average latency improved, despite the growing in the number of bigger packets that might happen in the network. Recall that the systems addressed in this paper have few very strict real-time communication flows, and a large number of less strict multimedia flows that tolerate much larger variations in latency and bandwidth. For those systems, the size of the real-time packets are related to control and signaling messages, and they are small when compared to multimedia packets. These are characteristics that we noticed on the industrial R&D projects carried out in the last decade.

### C. Offered load and latency

Remind that the equation 2 is based on the assumption of absence of contention. This Subsection introduces an evaluation of that equation for two hypothetical networks: one network with best-effort support and another one based on interleave of flits. We have chosen a best-effort for comparison because resources reservation results in poor utilization for applications that require variable-bit-rate communications [1], such the systems we are addressing in this paper.

Both networks were evaluated considering several values of offered traffic. For best-effort networks, the communication channels are shared by several flows. We have adopted as assumption that the best-effort NoC was implemented with a round-robin arbiter and wormhole switching. The latency for a flow $\sigma_i$ was calculated according the following equation:

$$L = (H_{path} \cdot t_r) + \frac{F}{|b-b_{occupied}|} \quad (4)$$

in which |b-$b_{occupied}$| is the bandwidth available for the flow under analysis $\sigma_i$, taking into account that $b_{occupied}$ of the whole bandwidth b has been used by other flows (offered traffic). Remember that for the wormhole switching, if a packet requests a communication channel that is being used by another packet, it must wait the another packet finish the transmission and release the resource communication channel before it starts its transmission.

The expression of latency for a hypothetical network based on the interleave of flits must take into account that there is no resources reservation in the network such happen for wormhole switching, and the flits from a packet may be interleaved with flits from other packets. Hence, the header latency $T_h$ must consider that N packets might compete at each router in the path, from its origin up to its destination $H_{path}$. It means that, under the latency point of view, the packet size grows N times. Thus, the expression of latency is given as follows:

$$L = (H_{path} \cdot t_r) + N \cdot (F/b) \quad (5)$$

A simulation of a SoC using these hypothetical networks was done based on the equations (4) and (5). It was done considering the following conditions:
- The number of routers in the path under analysis, for both networks, is 4 ;
- The router delay is the same for both networks, with the value 3; and
- There are three packets requesting the same resources for both networks. One of them, called $\sigma_i$, is the packet under analysis with fixed size of 100 flits, and the other two packets have variable sizes, from 0 up to 64k flits.

Figure 6 depicts the results generated by the simulation with equation (4) in gray color, and Equation (5) in black color. As expected, the latency for $\sigma_i$ was the same, when no other flows request the same resources (offered traffic = 0). With other two flows competing for the same resources, the latency for $\sigma_i$ in the NoC based on interleave of flits grows nearly three times; however, it keeps constant up to the maximum bandwidth usage. On the other hand, the latency of $\sigma_i$ in the best-effort network has grown when the offered load to the network is nearly 70%, as a result of network congestion for that flow under analysis. The result for interleave network was expected because the latency depends on the number of flits of the flow under analysis, and the number of flows that request the same resources, as shown in Equation (5).

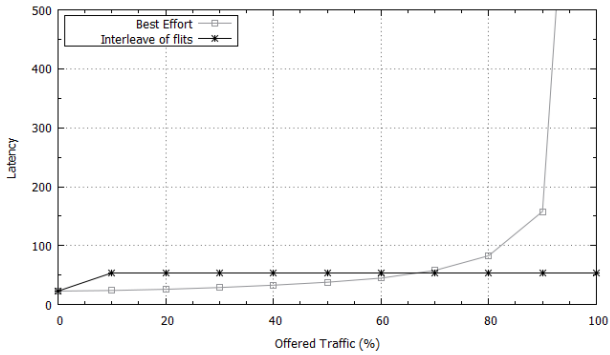

Figure 7. Simulation of interleaving for hypothetical BE network and the network based on interleaving of flits.

Similar result was introduced by [1], as shown Figure 7. The Figure shows network latency of a best-effort connection as a function of offered traffic measured for a single connection in a Æthereal NoC. The graph shows that latency is small and almost constant up to a certain turning point after which the latency grows steeply. This point is nearly at 75% of the offered load, before saturate. In that example, the latency saturates at 2600 ns because queuing between processing units and network interfaces is not taken into account by those authors.

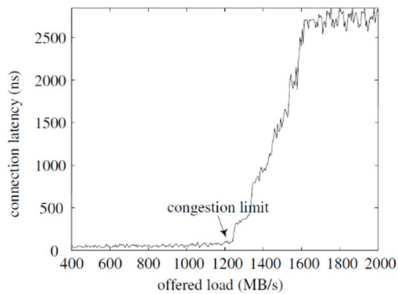

Figure 8. Network latency of an Æthereal BE connection as a function of offered load. Reference: [1].

Based on the information exposed in this Subsection, we understand that the method of interleave of flits can reduce the average latency under high traffic. The next step is to find out the worst-case latency (WCL) for the proposed network, what we have done and introduced in the next Subsection.

### D. Worst-case latency in Network-on-Chip

This Section introduces the additional latency that a NoC contributes to the execution time of the program instructions. Essentially, there are four distinguished types of network accesses in terms of the amount of data to be transferred: read/write of single-word transactions or read/write of block transactions. According [18], the execution time of a transaction involving a network access includes the time spent traversing the network ($T_{NoC}$) and the time spent accessing the remote core ($T_{core}$):

$$T_{transaction} = T_{NoC} + T_{core} \quad (4)$$

$T_{core}$ depends on the characteristics of the cores connected in the network. A transaction through the network may include up to three different delays. One delay is related to the time waiting before getting access to the network ($T_{wait;req}$). Another delay is related to the transaction request sent through the network ($T_{req}$). Finally, a reply should be send back, depending on the case, that would also require some waiting time for gaining access to the network ($T_{wait;reply}$) and some time to transfer the reply through network ($T_{reply}$), back to the requesting node. In general, the contribution of the network to the latency of a transaction may be given by:

$$T_{NoC} = T_{wait;req} + T_{req} + T_{wait;reply} + T_{reply} \quad (5)$$

Equation (5) is a general expression and may be adapted according the type of transaction in the network. The worst-case latency (WCL) for a packet that belongs to a flow $\sigma_i$ in the proposed network is defined as the sum of the latency experienced by all flits that belong to the same packet, on the path between an origin node and destination node with $h$ routers.

Our analysis for the WCL for packets in the network is based on the equation (2). The packet latency can be separated into two components: the time required for the header of the packet to traverse the network and the time for a packet of length $F$ to cross the channel with bandwidth $b$. The first flit of a packet in the proposed NoC is the header flit, and it might has a different latency than other flits of the same packet. It happens because once the first flit reach the destination node, than the other flits subsequent to it will be routing with the priorities established by the first flit in the path. Thereby, if there are no changes on other flows, then the payload and tail flits will have the same latency, that can be different from header flit.

The first latency to be analyzed in this Section is related to header flit. First, let us use as reference the bandwidth $b$ with one flit per clock cycle. Second, the latency to forward a flit from an input channel to an output channel is two clock cycles in the NoC's router we have proposed, as explained in

Section 0. If `N` flows are competing for the same output communication channel in the router, then all of these requests will receive permission to send a flit at each `N` arbitration cycle. Hence, the maximum latency expected for the header flit, $L_{header}$ that belongs to a packet from the flow $\sigma_i$ is given by the following expression:

$$L_{header} = \sum_{i=1}^{Hpath}(N_i \cdot t_r) = \sum_{i=1}^{Hpath} 2N_i \qquad (6)$$

Let's take into account the following assumptions: (*i*) the payload flits will be routing with the same priorities established by the header flit on routers in the path between the origin node and destination node, and (*ii*) all of the flows than might compete for the same resources are sending their packets to the same destination. Hence, the latency of payload and tail flit is given as following:

$$L_{payload;tail} = \frac{k(f-1)}{1/2} = 2k(f-1) \qquad (7)$$

in which `k` is the number of packets from other nodes in the whole network that are competing for the same destination node in the network and `f` is the amount of flits of the analyzed packet. From the equations (6) and (7), is possible to find out the worst-case latency for any packet in the proposed network, as following:

$$W_{packet} = \sum_{i=1}^{Hpath} 2N_i + 2k(f-1) + 2B \qquad (8)$$

in which `B` is the buffer size at network interfaces (FIFO memories). The buffer size `B` was multiplied by two because we are considering that is possible that both memories might be not empty with other flits, even in the origin node interface as in the destination node interface.

Note that the parameters in equation (8) are well known. Recall that the `XY` algorithm, implemented in the routers, is a static algorithm and imposes all flits that belong to the same packet must be routing by a unique path. Due to this algorithm's pattern, the maximum value of `N` will always be the same because the number of cores and routers in the path are fixed. The parameter `k` is also known due to the size of the network, and hence is possible to presume the maximum number of flows from other nodes in the whole network that might compete for the same destination node. Furthermore, the parameter `B` is defined at compilation time and the parameter `f` is known by origin node. It means that the real-time flows designed considering the absolute WCL of the proposed NoC will always meet the requirements of the associated real-time tasks, and hence there is no deadline lost.

## IV. EXPERIMENTAL RESULTS

In this Section, we introduce the experimental results based on experiments done with the NoC proposed in this paper, and it was organized in two Subsections. The first Subsection presents a hardware cost analysis based on silicon consumption for different configurations of the NoC using FPGA technology; meanwhile the other one introduces the result of hardware cost when the NoC is synthesized on ASIC technology.

### A. Hardware cost analysis on FPGA technology

We understand that there is a trade-off when adding destination information to the flits and more channels on the router, increasing silicon consumption. In this Subsection, we introduce the results related to the analysis that were done considering the silicon cost of the proposed NoC when synthesized for FPGA technology.

We synthesized instances of the NoC on a Xilinx Virtex 6 FPGA (XC6VLX75T-FF484) using Xilinx ISE 13.1. The NoC' router were synthesized using different network sizes and different number of communication channels. The results were given in terms of Slice Registers and Slice LUTs, and the following Subsections introduce these results making the point for this trade-off.

#### 1) Synthesis for different number of channels

In order to understand the impact caused by the number of channels per router, let us consider two hypothetical networks interconnecting twenty cores, each one. One of the networks is based on proposed routers with up to eight communication channels, shown in Figure 9-a, and the another one based on routers with up to five communication channels, shown in Figure 9-b.

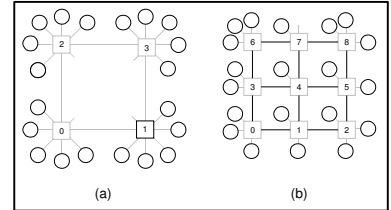

Figure 9. Hypothetical networks used as references to the silicon cost evaluations.

We synthesized two routers with 32 bits of data field: one with eight communication channels and another with five communication channels. We coarsely estimate the silicon consumption for the networks shown in Figure 9 multiplying the silicon resources of each router per number of router in those networks. The values of silicon consumption per router were obtained adding the consumption of Slice Register to the consumption of Slice LUTs, for each configuration. The results per router and network are introduced in Table I.

TABLE I. ESTIMATED VALUES OF SILICON CONSUMPTION FOR DIFFERENT CONFIGURATIONS OF NETWORKS.

| Channels/router | Consumption/router | Consumption/network |
|---|---|---|
| 5 channels | 2235 | 20115 (9 routers) |
| 8 channels | 3576 | 14304 (4 routers) |

The results from the Table I show that a network implemented using routers with eight channels per router has a silicon consumption nearly 29% lower than a network using routers with five channels per router. These values do not take into account the resources due to the implementation of communication channels that probably are bigger for the network using routers with five communication channels.

*2) Flit overhead analysis*

The proposed NoC has several design-time configurable parameters such as flit size, packet size, and number of channels. To evaluate the silicon resources and flit overhead trade-offs, we vary the parameters and report the results. We vary the number of extra bits needed to achieve different network sizes (i.e. number of extra bits per address field), and the data widths for communication (16, 32, 64, 128 and 256 bits). We define resources as the values of silicon consumption per component that were obtained adding the consumption of Slice Register to the consumption of Slice LUTs, and Table II summarizes the synthesis results.

TABLE II. FPGA RESOURCE CONSUMPTION CONSIDERING DIFFERENT DATA FIELD SIZES AND NETWORK SIZES.

| N. of extra bits | Slices Register | Slice LUTs | % Usage |
|---|---|---|---|
| *Data with 16 bits* | | | |
| 1 | 850 | 1821 | 1.9% |
| 2 | 894 | 1889 | 2.0% |
| 3 | 937 | 1969 | 2.1% |
| 4 | 979 | 2021 | 2.1% |
| *Data with 32 bits* | | | |
| 1 | 991 | 2070 | 2.2% |
| 2 | 1031 | 2145 | 2.3% |
| 3 | 1074 | 2225 | 2.4% |
| 4 | 1118 | 2227 | 2.4% |
| *Data with 64 bits* | | | |
| 1 | 1262 | 2589 | 2.8% |
| 2 | 1302 | 2667 | 2.8% |
| 3 | 1344 | 2737 | 2.9% |
| 4 | 1381 | 2789 | 3.0% |
| *Data with 128 bits* | | | |
| 1 | 1776 | 3613 | 3.9% |
| 2 | 1825 | 3681 | 3.9% |
| 3 | 1864 | 3761 | 4.0% |
| 4 | 1908 | 3813 | 4.1% |
| *Data with 256 bits* | | | |
| 1 | 2815 | 5661 | 6.1% |
| 2 | 2857 | 5729 | 6.1% |
| 3 | 2895 | 5809 | 6.2% |
| 4 | 2938 | 5861 | 6.3% |

Note that the increasing of extra bits used to address mesh networks from 2x2 up to 16x16 is not significant for the same Data field size. The silicon consumption is more relevant when the Data field size grows. For example, the increase of silicon consumption due to the growing number of bits used to address the routers in the network was in average 0.24%; meanwhile the consumption grows in average up to 4.18% when the Data field size changes from 16 bits to 256 bits.

*3) Hardware cost analysis on ASIC technology*

In this Subsection, we introduce the results related to the analysis that were done considering the silicon cost of the router we proposed when synthesized for ASIC technology. We use the Synopsys Ultra Design Compiler using SAED 90 nm Low Power technology, and the results were obtained in terms of silicon area for a single router, considering different size of communication channels (number of bits). We found out two NoCs also implemented using 90 nm silicon technology, and we compared those results with our router results.

The synthesis results for a 90 nm technology is shown in Table III. The router were implemented considering eight different channel sizes, starting with 32 bits and finishing with 256 bits. We chose this range of bits because it covers since single 32 bit processor up to GPUs. The last column of Table III shows, coarsely, the number of logic gates used to implement the router in 90 nm technology. The number of logic gates was given by the relation between the component area and the area of a basic gate (smaller) used by the target technology. For the SAED 90 nm library, it is a NAND gate with 5.5296 mm$^2$. This reference is useful to get an estimation of silicon area in other technologies, such as 65 nm for instance.

Further the area consumption for each router component, the Table III shows the contribution (%) of each component on the total area used by the router. Note that only the input interface, output interface and crossbar have a strong influence in the area consumption when the number with the growing of channel bits.

TABLE III. SILICON AREA (IN MM$^2$) USED BY A SINGLE ROUTER AND CONSIDERING DIERENT CHANNEL SIZES.

| Component | Channel size | | | |
|---|---|---|---|---|
| | *32 bits* | *64 bits* | *128 bits* | *256 bits* |
| Input interface | 0.0090 (9.44%) | 0.0150 (11.95%) | 0.0282 (14.48%) | 0.0541 (17.13%) |
| Output interface | 0.0021 (2.25%) | 0.0035 ( 2.77%) | 0.0062 ( 3.18%) | 0.0115 ( 3.64%) |
| Flow control | 0.0024 (2.47%) | 0.0024 ( 1.88%) | 0.0024 ( 1.21%) | 0.0024 ( 0.75%) |
| MUX | 0.0028 (2.96%) | 0.0049 ( 3.93%) | 0.0091 ( 4.69%) | 0.0171 ( 5.41%) |
| Allocator | 0.0013 (1.32%) | 0.0013 ( 1.01%) | 0.0013 ( 0.65%) | 0.0013 ( 0.40%) |
| Routing control | 0.0102 (0.35%) | 0.0102 ( 0.27%) | 0.0102 ( 0.18%) | 0.0102 ( 0.11%) |
| Arbiter control | 0.0003 (0.35%) | 0.0003 ( 0.27%) | 0.0003 ( 0.18%) | 0.0003 ( 0.11%) |
| Total area | 0.0954 | 0.1252 | 0.1946 | 0.3157 |
| Logic gates | 0.0172 | 0.0226 | 0.0352 | 0.0571 |

We also compare the cost of the router to the cost of other routers reported in the literature. A conclusive comparison is difficult to perform due to different features supported by other NoCs. Furthermore, most publications only report the cost of the routers without the cost of the interfaces used to connect the cores to the network. Nevertheless, we present a comparison using the available data synthesized for 90 nm.

Table IV shown the results, where we report the area after synthesis and compared to other values of router area reported in the literature that used the same 90 nm technology. The proposed router with 32 bits of payload occupies less area than the router proposed by [19] with 16 bits. Even when compared with dAElite, a network recognized in literature as a low cost NoC, our router with 64 bits of payload has consumption that we believe is acceptable, taking into account the result for 16 bits of [19].

TABLE IV. AREA COST OF PROPOSED ROUTER COMPARED TO OTHER 90 NM IMPLEMENTATIONS.

| NoC router | Characteristics | Area (mm$^2$) |
|---|---|---|
| Banerjee *et al* [19] | 4 SDM lanes 16 bits/lane | 0.108 |
| Proposed router | 32 bits (payload) and 8 ports | 0.095 |
| dAElite [9] | 64 bits/links divided in 4 TDM slots | 0.016 |
| Proposed router | 64 bit (payload) and 8 ports | 0.125 |

These Subsections above have introduced discussions about silicon cost, due to the adoption of a flit structure with routing information attached to all flits and the number of communication channels per router. As expected, our network has an increase on silicon cost caused by the extra bits on every flit. However, the adoption of routers without buffer memory minimized the growing, even for networks composed by routers with eight communication channels, despite it remains a penalty.

## V. CONCLUSION

This paper presented a Network-on-Chip with worst-case latency predictable at design time, in which there is no resources reservation. The proposed NoC architecture has based on the interleaving of flits from different flows in the same communication channel between routers, so each flit carries along routing information.

As expected, those routing information had increased the silicon consumption. However, the consumption is more relevant when the Data field size grows, than extra bits added in the flits. For example, an increase of silicon consumption due to the growing number of bits used to address the routers in the network was in average 0.3%; meanwhile the consumption grows in average up to 4.0% when the Data field size of the flit changes from 16 bits to 256 bits.

Despite the growing on silicon consumption caused by the adopted strategy, result depicted in Section III demonstrates that the average latency has kept lower the WCL boundary when the offered traffic is higher than 80%. This behaviour does not happen on regular BE schemes.

We also analytically demonstrate in Section III that real-time flows designed considering the absolute WCL of the proposed network will always meet the requirements of flows associated to real-time tasks, so no deadline can be lost due to network contention.